\documentclass[rnote]{aa} 

\usepackage{graphicx}
\newcommand{\water}{$\mathrm{H_2O}$}
\newcommand{\methanol}{$\mathrm{CH_3OH}$}
\newcommand{\hydroxyl}{$\mathrm{OH}$}
\newcommand{\kmps}{$\mathrm{km\,s^{-1}}$}
\newcommand{\smpy}{$\mathrm{M_\sun\,yr^{-1}}$}
\usepackage{txfonts}
\begin{document}

   \title{Periodic methanol masers in G9.62+0.20E}


   \author{D.J. van der Walt\inst{1}\thanks{email: johan.vanderwalt@nwu.ac.za} \and
     J.P. Maswanganye \inst{1,2} \and S. Etoka\inst{3} \and S. Goedhart \inst{4,1}  \and
     S.P. van den Heever \inst{1} }

   \institute{Centre for Space Research, North-West University, Potchefstroom, South
     Africa 
\and 
Hartebeesthoek Radio Astronomy Observatory, Krugersdorp, South Africa 
\and
Hamburger Sternwarte, Universit\"at Hamburg, Germany
\and 
SKA SA, The Park, Park Road, Pinelands, South Africa
}

   \date{}

 
\abstract{A number of mechanisms for understanding the periodic class II methanol masers
  associated with some high-mass star forming regions have been proposed in the past. Two
  recent proposals by \citet{parfenov2014} and \citet{sanna2015} have been presented in
  order to explain the periodic masers in sources with light curves similar to the
methanol masers in G9.62+0.20E. We evaluate to what extent the proposals and models
presented by these authors can explain the light curve of the methanol masers in
G9.62+0.20E. It is argued that neither of the proposed mechanisms can reproduce the light
curves of the methanol masers in G9.62+0.20E.}



   \maketitle
%

\section{Introduction}

Recently, two papers \citep{parfenov2014,sanna2015} have appeared in which explanations for the
periodic behaviour of class II methanol masers in the high-mass star forming region
G9.62+0.20E are presented.  \citet{parfenov2014} proposes that the flaring behaviour of
periodic masers with light curves such as in G9.62+0.20E might be due to rotating spiral
shocks in the gaps of circumbinary accretion disks around young binary stars. On the other
hand, based on the results of a high-resolution astrometric study of \methanol{},
\water{}, and \hydroxyl{} masers, as well as 7mm continuum emission of G9.62+0.20E,
\citet{sanna2015} suggest that the periodic masers in this source can be accounted for by
the presence of an independent pulsating young massive star within the context as proposed
by \citet{inayoshi2013b}. In addition to these two papers, \citet{szymczak2015} recently
presented the discovery of four new periodic masers and an updated light curve for
G22.357+0.066 which is strikingly similar to that of the 12.2 GHz masers in
G9.62+0.20E. These authors argue that although the colliding-wind binary (CWB) model of
\citet{vanderwalt2011} can explain the flare profile of G22.357+0.066, models in which the
maser flares are related to changes in the maser optical depths and excitation
temperatures -- such as that of \citet{inayoshi2013b} and \citet{parfenov2014} -- are 
preferred to the CWB explanation.

Given that the flaring behaviour of the 12.2 GHz masers in G9.62+0.20E and the 6.7 GHz
masers in G22.357+0.066 are so similar, it can be expected that the physical mechanism
underlying the maser flaring is the same in these two sources. Inspection of the light
curves presented by \citet{szymczak2015} rather strongly {\it suggests} that the flaring
behaviour of the masers in G45.473+0.134 has the same characteristics of that seen in
G9.62+0.20E and G22.357+0.066. Thus, at least for these three sources, we are
  faced with the question: Given the current available data, which of the proposed
  mechanims, i.e. a colliding-wind binary, a pulsating young high-mass star, or rotating
  spiral shocks in a disk seen edge on \citep[according to][]{parfenov2014}, is {\it the
    most probable mechanism} to explain the maser flaring?

In this note we argue on the basis of the expected maser flare profiles associated
  with a pulsating star driven by the $\kappa$ mechanism, that the proposal by
  \citet{sanna2015} cannot explain the observed light curves in G9.62+0.20E, and
therefore, by implication, nor  those of G22.357+0.066 and G45.473+0.134. We also
present an analysis of the model of \citet{parfenov2014} and argue that this model cannot
explain the flaring behaviour of the masers in the three sources mentioned.

\section{Analysis of models and proposals}

\subsection{ Proposal by \citet{sanna2015}}
The motivation for the proposal by \citet{sanna2015} that the periodic flaring of the
class II methanol masers in G9.62+0.20E is due to a pulsating young high-mass star is the
presence of a weaker radio continuum source 1300 AU from the strongest radio continuum
component. These authors argue that the stronger radio continuum source is most likely the
primary source of the infrared pumping of the \methanol{} masers. \citet{sanna2015}
consider that the weaker radio continuum source might be a bloated pulsating high-mass
protostar which provides the variable infrared emission that underlies the periodic
behaviour of the masers. For the pulsating high-mass protostar, \citet{sanna2015} assume
the model of \citet{inayoshi2013b}. In this model the protostar becomes pulsationally
unstable when the stellar radius expands maximally at a given accretion rate;  the
instability is caused by the $\kappa$ mechanism in the $\mathrm{He^+}$ layer. The pulsational
unstable state continues until the Kelvin-Helmholtz contraction stage is reached when the
stellar surface temperature increases and the $\mathrm{He^+}$ ionization layer disappears.
  \begin{figure}[h]
   \centering \includegraphics[width=9cm]{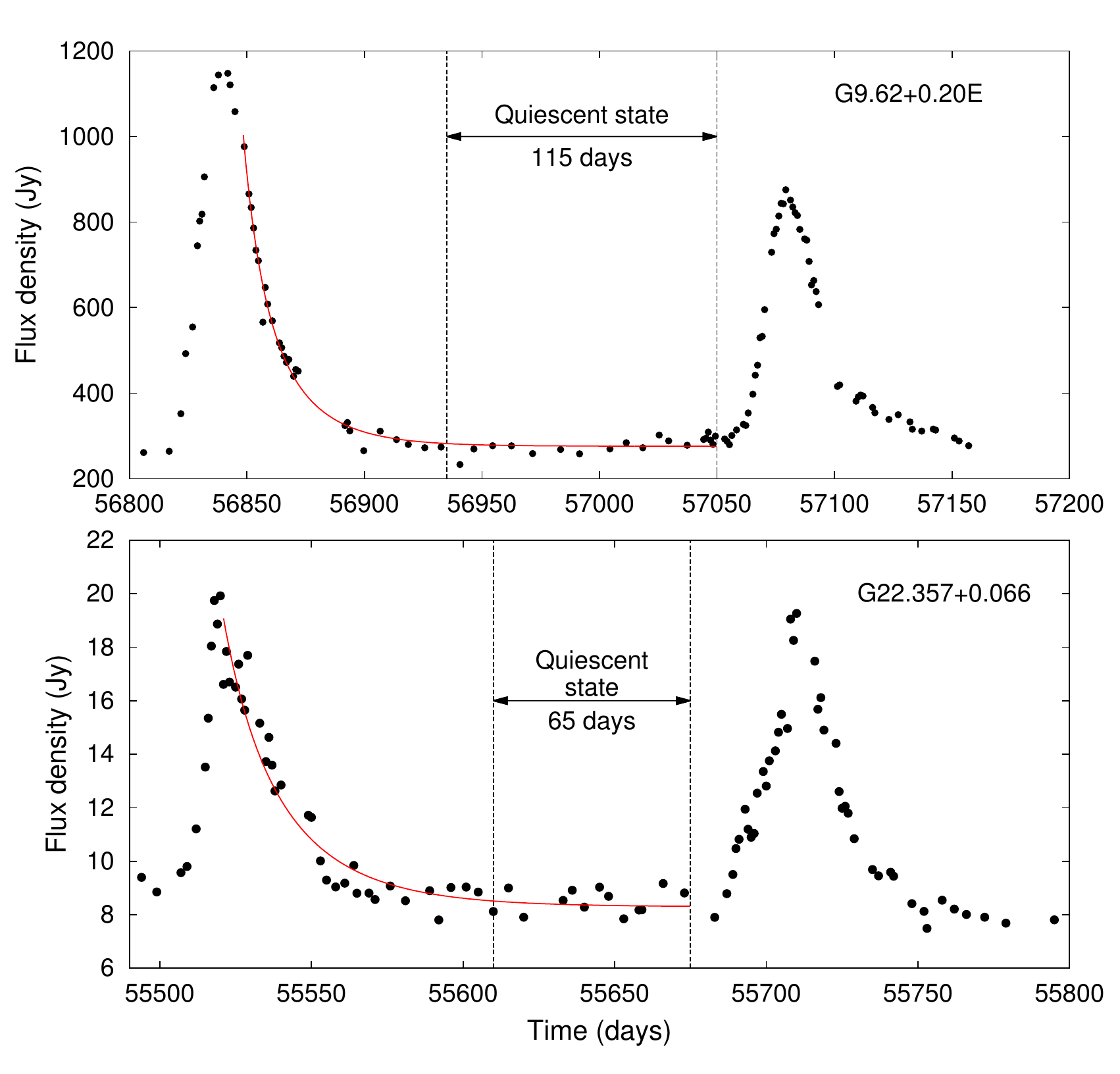}
   \caption{Selected parts of the timeseries for the 1.25 \kmps{} feature at 12.2 GHz for
     G9.62+0.20E (top panel) and the 80.09 \kmps{} at 6.7 GHz for G22.357+0.066. The red
     solid line in each of the panels is a fit to the decay parts of eq. A7 of
     \citet{vanderwalt2009}. The interval of the quiescent state was estimated by eye.}
   \label{fig:tsfit}
  \end{figure}

The important point to note here is that within the framework of the model of
\citet{inayoshi2013b}, the pulsations are driven by the $\kappa$ mechanism, similar to
that of other pulsating stars such as  the Cepheids. Thus, considering the physics
underlying the pulsations, it is reasonable to expect that the luminosity of the young
pulsating star should have a light curve similar to that of the Cepheids. 

The question now is what the expected light curve for the masers will be if the luminosity
of the pulsating star, which heats the surrounding dust and therefore affect the pumping
of the masers, varies in a way similar to that of the Cepheids. To address this question
we use the OH masers associated with Mira variables as examples of pulsating stars for
which both the optical and maser light curves are known and from which the relation
between the OH maser and optical light curves can be deduced. In this regard,
\citet{etoka2000} have presented and compared the behaviour of the OH masers and optical
light curves for seven Mira variables. These authors conclude that although the optical
light curves are more strongly asymmetric than that of the associated OH masers, the
shapes of the optical and OH maser light curves are very similar. This applies to 1612
MHz, 1665 MHz, and 1667 MHz maser transitions even though the degree of saturation does
not seem to be the same for all three transitions. Although the example of the Miras might
not be exactly equivalent to the scenario envisaged in the proposal of \citet{sanna2015},
it nevertheless strongly suggests that if the optical light curve of the pulsating star in
the model of \citet{inayoshi2013b} is similar to that of the Cepheids, then it can be
expected that the light curve of the methanol masers should be similar in general.

One of the characteristics of the light curves of the Cepheids is that it {\it never}
reaches a quiescent state, that is, a state in which the star's luminosity stays constant
for a significant fraction of the period \citep[see e.g.][]{yoachim2009}. This is also the
case for OH masers associated with Miras \citep{etoka2000} as well as for the OH masers
associated with OH/IR stars \citep{engels2015}. However, for both the 12.2 GHz
masers in G9.62+0.20E and the 6.7 GHz masers in G22.357+0.066, a well-defined quiescent
state can be identified (see Fig. \ref{fig:tsfit} and \citet{szymczak2015}). In the case
of G9.62+0.20E, the quiescent state lasts for about 115 days (47\% of the period) and for
G22.357+0.066 it is about 65 days (37\% of the period). Should the flaring of the methanol
masers in G9.62+0.20E be due to the effect of a pulsating star driven by the $\kappa$
mechanism, then the quiescent state of the maser would imply that the star has settled
into an equilibrium state for a significant fraction of the period. The physics of the
$\kappa$ mechanism, however, does not allow the star to reach equilibrium after each flare
and to spontaneously start to pulsate again. Within the framework of the model of
\citet{inayoshi2013b}, it would mean that after a flare the $\mathrm{He^+}$ ionization
layer disappears but spontaneously reappears at the end of the quiescent phase at which
time there should also be a mechanism that trigger the pulsation of the star. 

Considering the above, we conclude that the behaviour of the periodic methanol masers in
G9.62+0.20E is inconsistent with what is expected if flaring of the masers is driven by a
pulsating young high-mass star within the framework of the model of \citet{inayoshi2013b}.

\subsection{ Model of \citet{parfenov2014}}

The \citet{parfenov2014} model is significantly more complicated than that of a pulsating
star; furthermore there is no direct analogue for this scenario similar to what the Mira
variables and OH/IR stars are for the proposal by \citet{inayoshi2013b}.  A longer
discussion is therefore needed. We first present a brief outline of the model of
\citet{parfenov2014}. Based on the information in the paper by \citet{parfenov2014}, we
constructed a geometric model which is used to calculate an expected light curve.  We also
comment on the properties of the bow shocks as used by \cite{parfenov2014} since it has
direct implications for the predictions of their model. Finally, we consider the effect
that the presence of a stellar wind associated with the more massive star of the binary
system might have on the locations and luminosities of the bow shocks. Such a
consideration is necessary since it also has direct implications for the predictions of
the model by \citet{parfenov2014}.

\subsubsection{Brief description of the model}
The ideas in the model of \citet{parfenov2014} are to a large extent based on the
scenarios described by, for example, \citet{ochi2005} and \citet{sytov2011}, although use
is also made of the results by \citet{sytov2009}. \citet{sytov2011},
\citet{kaigorodov2010}, and \citet{fateeva2011} have run numerical simulations to determine the
structure of circumbinary envelopes and the flow of gas in young binary systems. The
binary systems considered by these authors consist of lower mass and/or T-Tauri
stars. These calculations show the creation of a central region of low density (referred to
as the central gap) through the action of bow shocks that form in front of the
circumstellar accretions disks of the two stars in the case when the orbital speed of the
stars is supersonic. Associated with the bow shocks are also trailing spiral shocks that
extend from the outer edges of the circumstellar disks to the inner edge of the
circumbinary disk.

  \begin{figure}[h]
   \centering \includegraphics[width=6.5cm]{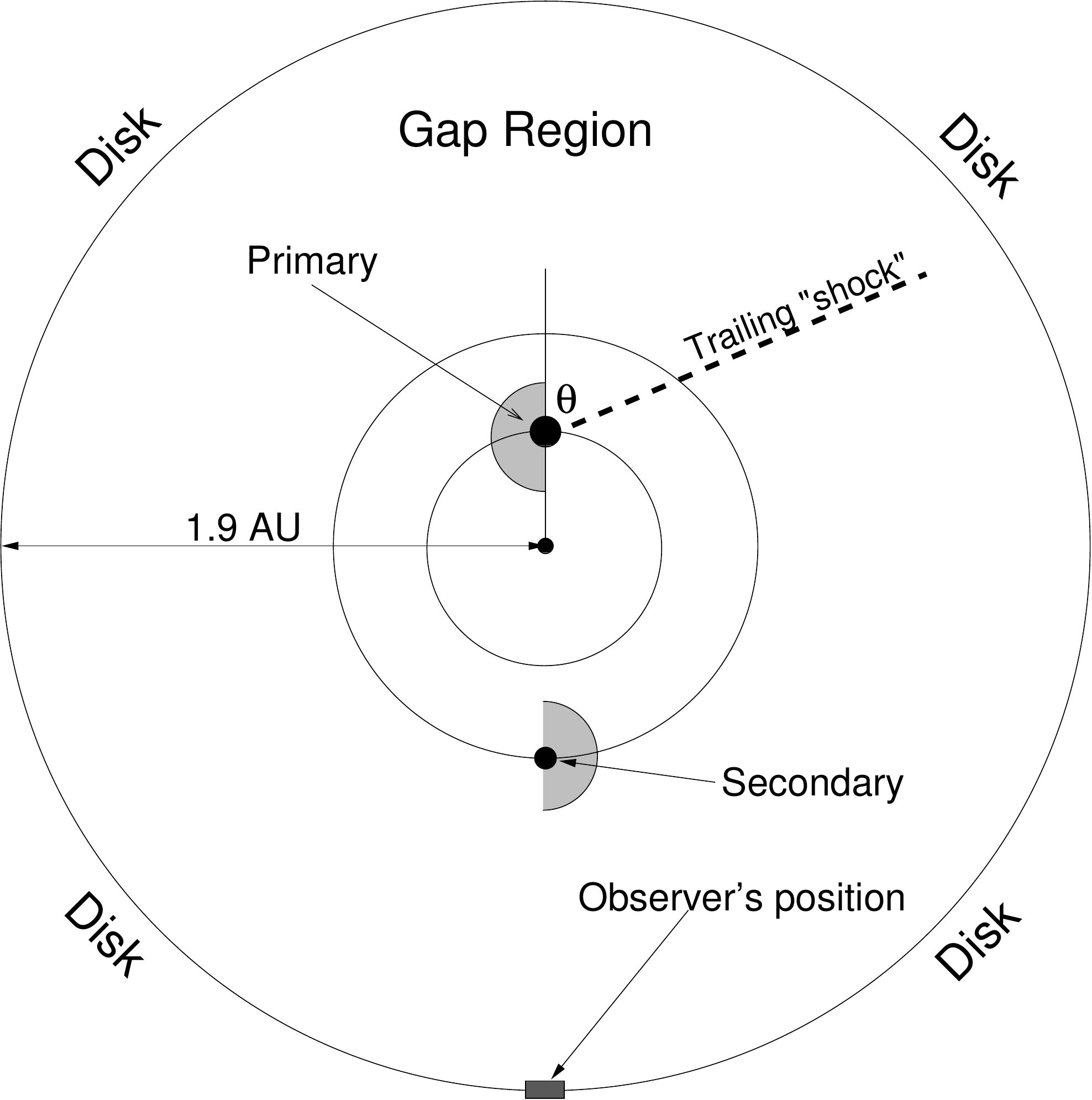}
   \caption{Schematic representation of the model of \citet{parfenov2014}. See text for
     more details. The orbital motion is in an anticlockwise direction.}
   \label{fig:geometry}
  \end{figure}

\citet{parfenov2014} transferred some of the elements of this  scenario into
their model.  The basic ingredient is also a binary system inside the gap region of a flat
circumbinary disk. The gap region is circular with a radius of 1.9 AU. The two stars have
masses of 13 M$_\sun$ (the primary) and 7 M$_\sun$ (the secondary), respectively. The
semimajor axis of the circular orbit is 1.145 AU with a period of 100 days.  Using the
masses  given above, the 13 M$_\sun$ star follows a circular orbit with radius 0.4 AU
around the centre  of mass while for the 7 M$_\sun$ star the radius of its orbit is 0.74
AU. The orbital velocity  is then found to be 43.6 \kmps{} for the primary star and 81
\kmps{} for the secondary star. \citet{parfenov2014} do not make reference to the orbital
motion of the secondary and it is not clear why, since both stars follow circular orbits
around the centre of mass. Within the description given by \citet{parfenov2014}, a bow
shock (to which they also refer as a spiral shock wave) is associated only with the
primary star.  The material behind the bow shock is dense and hot, resulting in strong UV
radiation that can heat the gas and dust on the inner edge of the circumbinary disk. The
rotation of this shock is considered to be the main cause for changes in the dust temperature
in the circumbinary disk and therefore for the periodic flaring of the masers 
located in the circumbinary disk. \citet{parfenov2014} speculates that the maser
brightness traces the change in gas column density along the line between the centre of
the disk and a point on the inner edge of the circumbinary disk. This change in gas column
density will then give rise to a rather sharp increase in the maser brightness followed by
a slow decay, thus explaining the observed light curve of the periodic methanol masers in
sources like G9.62+0.20E. In their  modelling, the shock is modelled as a half disk with an
inner radius of 0.022 AU (the radius of the primary) and an outer radius of 0.2 AU.

\subsubsection{Simple geometric model}

Following the description given by \citet{parfenov2014}, we constructed a schematic
representation of their model (Fig. \ref{fig:geometry}). \citet{parfenov2014} use the
terms ``bow shock'' and ``spiral shock'' interchangeably. Here we use the term bow shock
only for the shock that forms ahead of the star.  We also represent bow shocks with half
disks, although in our case the leading edges of the bow shocks are in the instantaneous
directions of the velocity vectors of the two stars and not perpendicular to the velocity
vector as shown in Fig. 1b of \citet{parfenov2014}.  As shown in Fig. \ref{fig:geometry},
a bow shock should be associated with the secondary since its orbital velocity is also
supersonic. In this regard, we point to the results of \citet{kaigorodov2010} where bow
shocks are associated with {\it both} stars and the bow shocks form {\it ahead} of the
circumstellar disks associated with the two stars. In what follows, we therefore use the
orientation of the bow shocks as shown in Fig. \ref{fig:geometry}. For simplicity, we
consider the trailing spiral shock to have the linear structure shown in
Fig. \ref{fig:geometry}. We note that, according to e.g. \citet{kaigorodov2010}, spiral
shocks are associated with the circumstellar accretion disks around the primary and the
secondary. \citet{parfenov2014} make no mention of such circumstellar accretion disks.

\subsubsection{Expected light curve}

\begin{figure}[h]
   \centering \includegraphics[width=9cm]{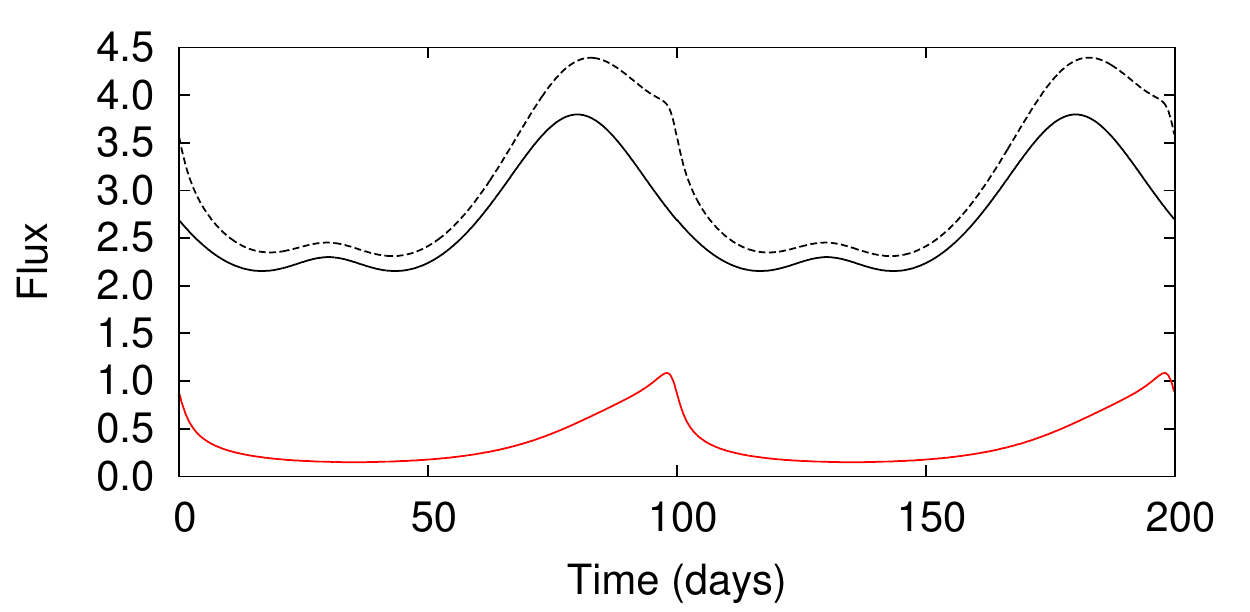}
   \caption{Expected time dependence of the illumination of a specific point on the inner
     edge of the circumbinary disk for the cases where the two stars only are considered
     (solid black line) and for the trailing shock (red line). The values of the
     parameters used for the trailing shock are given in the text. The dashed black
     line is the sum of the two cases. }
   \label{fig:binpulse}
  \end{figure}

We first want to remark on the general shape of the light curve expected from such a
binary system. The underlying dynamical mechanism in this scenario is the binary system.
The orbital motion of the stars, and any other emitting structures associated with the
stars, modulates the radiation that illuminates a particular point on the inner edge of the
circumbinary disk (henceforth referred to as the observer). The modulation is purely
geometric due to the time-dependent variation of $1/d^2$, where $d$ is the distance
between each of the stars and the observer. Considering only the two stars (taken as point
sources) it is straightforward to calculate the time dependence of the radiative energy
flux at the position of the observer given that the ratio of the luminosity of the primary
to the secondary is 8.2:1.  No optical depth effects were taken into account. The result
is shown in Fig. \ref{fig:binpulse} as a solid black line. The larger peak is due to the
primary star. The effect of the secondary is significantly less because  its luminosity
is only about 12\% of that of the primary.

It is also rather simple to calculate the effect of the spiral shock if it is approximated
as a trailing linear structure as shown in Fig. \ref{fig:geometry}. It can be expected
that the volume emissivity of the gas decreases in some way along the length of the
shock. To account for this behaviour, we assumed the volume emissivity to decrease as
$(r/r_p)^{-q}$, where $r$ is the radial distance from the disk centre of a point on the
shock and $r_p$ is the orbital radius of the primary star. The spiral shock is then
considered to consist of a large number of point sources rotating around the centre of
mass. The total luminosity of the shocked gas is set to some fraction, $\alpha$, of the
luminosity, $L_p$, of the primary; the luminosity of the $i$-th point source is then $L_i
= L_0 (r_i/r_p)^{-q}$. The value of $L_0$ is determined from the condition that $\sum_i
L_i = \alpha L_p$.

As an {example} to illustrate the effect of the linear trailing shock,  in
Fig. \ref{fig:binpulse} we
show the time dependence of the illumination at the position of the
observer for the case when $\theta = 70^{\circ}$ (see Fig. \ref{fig:geometry}), the end
point of the shock has an orbital radius of 1.71 AU, $q = 2 $, and the total radiative
luminosity of the shock is equal to the luminosity of the secondary star. In reality, the
radiative luminosity of the shocked gas may be significantly lower. It is important to note
the shape of the light curve: starting from the first minimum there is a rather slow
increase in the flux up to a maximum, followed by a rapid decrease towards a minimum. The
slow increase occurs when the shock approaches the observer; the
different parts of the shock are at different distances, $d$, from the observer and 
the maximum contribution from the different parts of the shock to the flux at the observer
occurs at different times. The maximum in the light curve occurs when the end point of the
shock is on its nearest position to the observer, i.e. on the line connecting the centre of
the circumbinary disk with the observer. At that position the distance between the base of
shock (using the terminology of \citet{parfenov2014}) and the observer is already so large
that the contribution by the base of the shock to the flux is not significant. As the
shock moves further, the flux decreases very rapidly owing to the rapid increase in the
distance between the end point of the trailing shock and the observer. It is also seen
that for the case considered, the peak in the flux due to the shock is delayed relative to
the peak due to the primary. The degree to which it is shifted depends on the angle
$\theta$ as defined in Fig. \ref{fig:geometry}.  Although we used a linear geometry for
the shock, the same asymmetric type of light curve is expected for a shock with a spiral
structure as in the models of e.g. \citet{kaigorodov2010}, \citet{sytov2011}, and
\citet{fateeva2011}.

We have not attempted to calculate the light curve at the position of the observer due to
the two bow shocks. Since the presence of the two bow shocks does not significantly break
the symmetry for the case when only the two stars are considered, it is expected that the
shape of the light curve with the bow shocks included will be rather similar to that of
the two stars only.  

Now, the case of a binary system, with or without bow or spiral shocks, as the variable
source of radiation that heats the dust on the inner edge of the circumbinary disk is in
principle no different from that of the Miras or the OH/IR stars discussed earlier.
Radiative transfer and reprocessing by dust of the radiation incident on the inner edge of
the circumbinary disk up to the region where the masers operate, as well as the response
of the masers, will definitely further modify the incident light curve to eventually
produce the observed light curve of the masers. Given the example of the Miras, it can be
concluded that if the light curve of the variable source at the position of the observer
is that shown by the dashed line in Fig. \ref{fig:binpulse}, then the masers should have a
similar light curve, although not exactly the same.  It is therefore difficult to see how
a combination of the orbital motion of the stars, the presence of bow or spiral shocks,
the normal dust reprocessing of radiation as well as the response of the masers can
produce a final maser light curve that resembles the light curve seen for example in  the 12.2 GHz
methanol masers in G9.62+0.20E (Fig. \ref{fig:tsfit}).

Our conclusion here is that the expected light curve of the masers in the binary scenario
proposed by \citet{parfenov2014} will be quite different from that suggested by these
authors and that the observed methanol maser flaring in G9.62+0.20E cannot be explained
with this model.

\subsubsection{Comments on the properties of the shocks in the model of \citet{parfenov2014}}

In the above analysis, we assumed the bow shocks to have the properties  used by
\citet{parfenov2014}. However, it is necessary to note the following: \citet{parfenov2014}
modelled the bow shock as a half disk of uniform temperature (30 222K) and density. Two
densities are assumed,  10$^{13.32}$ and 10$^{13.62}$ cm$^{-3}$. The maximum jump in
density across a shock is a factor of four in the case of a strong shock
\citep{dopita2003,lequeux2005}. Assuming the strong shock limit to apply, the implication
is that the density in the preshocked gas had to be respectively 10$^{12.72}$ and
10$^{13.02}$ cm$^{-3}$ for the above assumed densities of the postshocked gas. However,
\citet{parfenov2014} assumed a hydrogen density of 10$^{9.8}$ cm$^{-3}$ at the inner edge
of the circumbinary disk, which is significantly lower than the value expected for the
preshocked gas assuming the strong shock limit. If the inner edge of the circumbinary disk
is to be well defined, then the density of the ``low-density hot gas'' in the gap region
should be lower than 10$^{9.8}$ cm$^{-3}$. Considering the results of
\citet{kaigorodov2010} and \citet{fateeva2011}, the density in the gap region is {\it at
  least} two orders of magnitude lower than at the inner edge of the circumbinary
disk. This would mean that the gas in the gap region should have a density of at most
10$^{7.8}$ cm$^{-3}$ and the postshocked gas in the bow shock a density of 10$^{8.4}$
cm$^{-3}$, i.e. at least five orders of magnitude lower than that used by
\citet{parfenov2014}. The implications of such a low value for the density of the
postshocked gas is obviously that the luminosity of the shocks will be significantly lower
than estimated by \citet{parfenov2014}.

\subsubsection{ Effect of stellar winds}

There is one more aspect related to bow shocks that should be considered. Although
\citet{parfenov2014} make no mention of stellar winds, it is reasonable to assume, from
the {\it given} properties of the stars, that both stars have stellar winds.  Using Fig. 4
of \citet{kudritzki2000}, the wind speed for a main-sequence star with T$_{eff}$ = 20 000
K is of the order of a few 100 \kmps.  In the case of a main-sequence star with T$_{eff}$
= 29 000 K, it ranges between about 1000 \kmps and about 2300 \kmps. For illustrative
purposes, we will use wind speeds of 800 \kmps and 1500 \kmps\ for the secondary and
primary stars, respectively. Also associated with the winds is the mass-loss rate,
$\dot{M}$. For this, we use the results of \citet{vink2000} and take $\dot{M}$ to be
$10^{-7.5}$ \smpy \ (their Fig. 2) for the primary. The parameters of the secondary fall
outside of the range covered by the calculations of \citet{vink2000}, but we will
arbitrarily take $\dot{M}$ as $10^{-8.5}$ \smpy.

Given the wind velocities and the mass-loss rates, it is now possible to calculate the
stand-off distances of the bow shocks and their luminosities due to the supersonic speed
of {\it both} stars and the interaction of their winds with the low-density gas in the gap
region. The stand-off distance is given by \citet{wilkin1996} as
\begin{equation}
R_0 = \sqrt{\frac{\dot{M}V_w}{4\pi \rho_a V_s^2}}
\label{eqn:r0}
,\end{equation}
where $\rho_a$ is the mass density of the low-density gas in the gap region and $V_w$ and
$V_s$ are respectively the wind and stellar velocities. Using for $\rho_a$ the value
corresponding to 10$^{7.8}\,\mathrm{cm^{-3}}$ , we find stand-off distances of 7.3 AU and
0.9 AU for the bow shocks associated with the primary and secondary, respectively. The
implication of the value of 7.3 AU for the stand-off distance for the bow shock associated
with the primary is that, rather than forming a bow shock, the wind of the primary will
directly interact with the inner edge of the circumbinary disk.

The {\it upper limit} for the radiative luminosity of a bow shock
is given by \citet{wilkin1997} as

\begin{equation}
L_{shock} = \frac{1}{2}\dot{M}(V_w^2 + V_s^2)
\label{eqn:lshock}
.\end{equation}

From this, we find maximum radiative luminosities of 5.92~L$_\sun$ and 0.17~L$_\sun$ for
the bow shocks associated with the primary and secondary respectively. Both of these are
significantly smaller that the luminosities of the associated stars. The implication of
these numbers is that the associated bow shocks do not play a significant role in the
radiative energy that hits the circumbinary disk's inner boundary and is in contradiction
with the results of \citet{parfenov2014}.

The presence of the winds also implies the interaction of the winds. This means that there is  a contact discontinuity between the two stars with shocked wind material on
both sides of it. With the wind speeds assumed above, and using equation 2 of
\citet{parfenov2014}, the temperature of the shocked wind of the primary will be about
$3.6\times 10^7$K and $10^7$K for the shocked wind of the secondary. The system proposed
by \citet{parfenov2014} is therefore actually a colliding-wind binary within the gap of a
circumbinary disk. If $d_1$ and $d_2$ respectively denote the distances to the contact
discontinuity of the primary and the secondary, then the ratio $d_2/d_1$ is given by
\begin{equation}
\frac{d_2}{d_1} = \sqrt{\frac{\dot{M}_2 V_{w,2}}{\dot{M}_1 V_{w,1}}}
\end{equation}
\citep{stevens1992}.
From this expression it is found that $d_1$ = 0.93 AU, which means that the contact
discontinuity is situated on the same side as the secondary and has a circular orbit with a
radius of 0.19 AU around the centre of mass. Although the temperatures of the shocked winds
are quite high, the radiative luminosity of the shocked gas is low. Using the
wind speed and mass-loss rate for the primary as above, the total mechanical luminosity of
the wind is 5.9~L$_\sun$. If, for example, gas flowing out in $\pi$ steradians are
effectively shocked, then the upper limit for the radiative luminosity of the shocked gas
of the primary is only about 1.48~L$_\sun$, which in this case is only 0.01\% of the
luminosity of the primary.

\section{Discussion and conclusions}
Although it is necessary to consider all available observational data when trying to
explain the periodic masers in a specific source, it cannot be denied that the light curve
is certainly one of the most important pieces of observational information that points to
the mechanism underlying the periodic behaviour. The case of the OH masers associated with
Mira variables is a real example which illustrates  that the light curves of the
OH masers reflect the basic underlying periodic mechanism. The same is expected to apply
to all periodic methanol masers. As already argued above, the light curves expected from a
pulsating star or from the binary system as proposed by \citet{parfenov2014} are completely
different from the observed light curve of the periodic methanol masers in G9.62+0.20E. It is
therefore unlikely that either of these two mechanisms underlie the periodic flaring of
the methanol masers in sources such as G9.62+0.20E. 

In the upper panel of Fig. \ref{fig:tsfit},  we also show the fit of equation A7 of
\citet{vanderwalt2009} to the decay part of the first of the two shown flares. Considering
the quality of the fit over an interval of 200 days, it seems reasonable to conclude that
the flaring of the methanol masers in G9.62+0.20E is likely to be due to the maser
reflecting changes in the flux of free-free emission from a background thermal hydrogen
plasma being excited from an equilibrium state of ionization (as set by the presence of
the high-mass star) to a higher state of ionization, followed by recombination to the
level set by the equilibrium state. There must, however, be a mechanism to explain the
periodic increase of the ionization level of that part of the \ion{H}{II} against which
the maser is projected. Whether  the underlying driving mechanism is a
colliding-wind binary system has to be verified with other observations, e.g. the presence
of associated X-ray emission.

To what extent there are other mechanisms that can equally well describe the maser flare
profile in G9.62+0.20E (and sources with similar maser flare profiles) is not clear since
periodic variability can only be associated either with the young star or with a binary
system. From our discussion above it seems, however, that different physical mechanisms
used to explain the periodicity will give rise to different flare profiles. Having ruled
out the possibility that it can be due to the pulsation of the young star, it follows that
the methanol maser flare profiles in G9.62+0.20E (and other sources with similar flare
profiles) are most probably due to some periodic process associated with a binary system
that affects the ionization level of parts of the background \ion{H}{II} region.

Although our discussion above is focused on the periodic masers in G9.62+0.20E, the
general requirement is that any proposed model for the periodic masers in a particular
source must also be able to explain the observed maser light curve. It was noted rather
explicitly by \citet{szymczak2015} that three of their newly detected periodic masers have
light curves consistent with the CWB model of \citet{vanderwalt2011}. Together with
G22.357+0.066, this brings the number of periodic masers with light curves similar to
G9.62+0.20E to five. Inspection of the light curves of the remaining nine periodic maser
sources shows that the majority has the characteristic of not showing a quiescent
state. Whether the periodicity in these sources is driven by a pulsating young star is
not clear. Since there is no explicit example of a pulsating young massive star it cannot
be excluded that there might be pulsation mechanisms other than that proposed by
\citet{inayoshi2013b}, which might be able to explain the light curves of periodic methanol
masers that cannot be explained within the framework of the colliding-wind binary model.

\bibliographystyle{aa}

\bibliography{ref}

\begin{acknowledgements}
This work was supported by the National Research Foundation under Grant number 2053475.
\end{acknowledgements}


\end{document}